\documentstyle[mprocl]{article}
\def\be{\begin{equation}}
\def\ee{\end{equation}}
\def\bea{\begin{eqnarray}}
\def\eea{\end{eqnarray}}

\begin{document}
\title{SELF-ORGANIZED CRITICALITY AND 1/f NOISE IN TRAFFIC}
\author{ MAYA PACZUSKI \footnote{Please address correspondences to
maya@cmt1.phy.bnl.gov.} }
\address{Physics Department, Brookhaven National Laboratory \\
         Upton, NY  11973, USA}
\author{ KAI NAGEL }
\address{Los Alamos National Laboratory, P.O. Box 1663, MS997 \\
        Los Alamos, NM  87545, USA }
\maketitle
\abstracts{Phantom traffic jams may emerge ``out of nowhere'' from
small fluctuations rather
than being triggered by large, exceptional events.  We show how phantom
 jams arise in a model of single lane highway traffic, which mimics
human driving behavior.
Surprisingly, the optimal
state of highest efficiency, with the largest throughput,
is a critical state with traffic jams of all sizes.  We demonstrate that
open systems self-organize to the most efficient state.
In the model we study, this critical state is a percolation transition
for the phantom traffic jams.  At criticality,
the individual jams have a complicated fractal structure where cars
follow an intermittent stop and go pattern.  We analytically derive
the form of the corresponding power spectrum to be  $1/f^{\alpha}$
 with $\alpha =1$ exactly.  This theoretical
prediction agrees with our numerical simulations
and  with observations of $1/f$ noise in real traffic.\cite{ np95}}
Content-Length: 29750
Content-Type: text
X-Lines: 696
Status: RO

\section{Introduction}
Our everyday experience with traffic jams is that they are annoying and
worth avoiding.  Intuitively, many people believe
 that if we could somehow get rid of
 jams then traffic would be more efficient with higher throughput.
Here we show that this is not necessarily true.  By studying a simple
model of highway traffic, we find that the state with the highest
throughput is a critical state with traffic jams of all sizes.  If
the density of cars were lower, the highway would be underutilized; on the
other hand, if it were higher there would inevitably
be a huge jam lowering throughput.
This leaves us with the critical state as the most efficient state that
can be achieved.

Finding a real traffic network operating at or near peak efficiency
may seem highly unlikely.  To the contrary, we find that an
open network self-organizes to the critical state.  The output from
large jams finds the maximum allowed throughput which is barely stable to
perturbations.  Small perturbations in the outflow lead to traffic
jams of all sizes.  These phantom traffic jams may be viewed as
avalanches which form a fractal in space and time.  A snapshot of the
jam at a certain point in time can be represented as a spatial
fractal.  A time series of the jam at a certain point in space will
also be a fractal in time with a $1/f$ power spectrum.\cite{ np95}
  The space and
time snapshots are related to each other since they are different cuts
of the same underlying avalanche.\cite{ mpb94}  This picture of avalanche
dynamics has application to many dynamical systems in addition
to traffic.\cite{ pmb96}

Actually about twenty years ago, it was discovered that traffic is an
example of $1/f$ noise.  T. Musha and H. Higuchi studied traffic on a
three lane section of the Tomei Expressway in Japan.\cite{ musha} They
recorded the transit times of cars passing underneath a bridge
spanning the motorway.  They discovered that the power spectral
density of the car current fluctuations had a $1/f$ low frequency
behavior.  This type of power spectrum is actually common to many
granular flow systems, in addition to traffic.\cite{ hh} We find that
one of the ramifications of criticality in traffic are long tails in
the temporal correlation functions.  We explicitly derive a $f^{-1}$
spectrum from the microscopic dynamics of a simple model for traffic.
It is possible that $1/f$ noise in other granular systems can be
understood from similar considerations.

In addition,
we show that the fluctuations found in the 5-min measurements of traffic
at capacity, by Hall  and others,\cite{ hall}
 reflect the fact that traffic flow is intermittent
and inhomogeneous with essentially
two states (jammed and maximum throughput).  
 We end our presentation highlighting applications
to real traffic and a view toward economic systems.

\section{The Model}

The Nagel-Schreckenberg~\cite{ ns92}
model that we study is defined on a one dimensional lattice with cars moving to
the right.  Cars can move with integer velocities in
the interval $[0,v_{max}]$. 
The maximum velocity $v_{max}$ is typically set equal to 5.
This velocity defines how many ``car lengths'' each car will move
at the next time step.  If a car is moving too fast, it must slow
down to avoid a crash.  A slow moving car
will accelerate, in a sluggish way, when given an opportunity.
The ability to accelerate is slower than the ability to
break.  Also, cars moving at maximum velocity may slow down
for no reason, with
probability $p_{free}$.
For more details of the model, see K. Nagel's article in the
same volume.

\subsection{Cruise Control Limit}
We consider a limit of the model where $p_{free} \rightarrow
0$.  This means that
all cars which have reached maximum velocity, and have enough headway
in front of them to avoid crashes, will continue to move at maximum
velocity.  Thus it is possible for the motion in the system to be
completely deterministic.

For every  configuration of the model, one iteration
consists of the following steps, which are each performed
simultaneously for all vehicles (here, the
quantity $gap $ equals the number of empty sites
in front of a vehicle):\begin{itemize}

\item 
A vehicle which travels at maximum velocity
$v_{max}$ and has free headway: $gap \ge v_{max}$
just maintains  its velocity.

\item
All other vehicles are jammed.
The following two rules are
applied to jammed vehicles:\begin{itemize}

\item
{\bf Acceleration of free vehicles:} With probability~1/2, a vehicle
with $gap \ge v+1$ accelerates to $v+1$, otherwise it keeps the
velocity~$v$.

\item
{\bf Slowing down due to other cars of car following:} Each vehicle
with $gap \le v$ slows down to $gap$: $v \to gap$.  With
probability~1/2, it overreacts and slows down even further: $v \to
\max[gap-1,0]$.

\end{itemize} 

\item {\bf Movement:} Each vehicle advances $v$ sites.  

\end{itemize}

\section{Phase Transition in a Closed System}

Before considering the behavior of an open system, it is worthwhile to
study a closed system, i.e. a ring, where the number of cars is held
fixed.  Our fundamental diagram, or current-density relation
$j(\rho)$, is indicated in Fig. 1.  These diagram involves
steady-state, long time averages over the entire
system starting from a random initial condition (see Ref. [1]).  At
low density, the steady state flow is laminar.  All the jams present
in the initial configuration have decayed away.  In other words, they
have been ``absorbed'' by the deterministic state where all cars move
at maximum velocity.  As a result the linear slope at low density is
just $v_{max}$.  At high density, jams present in the initial
configuration are never erased from the system.  These long lived jams
lower the throughput and the current decreases linearly with
increasing density.  So the maximum throughput point separates the
deterministic state with no jams at long times from the state with a
finite density of jammed cars.  It corresponds to a continuous phase
transition where the order parameter, $m$, is the density of jammed
vehicles.

Even though the deterministic state is not dynamically accessible
above maximum throughput, $j_{max}$, it is possible to prepare special initial
configurations that have no jams.  In this case, the steady state will
also have no jams, and the current will still be a linearly increasing
function of $\rho$ (the dotted line in Fig. 1.).  In this sense, the
high density phase is bistable.  But  perturbations of the high
density deterministic branch will lead to long-lived traffic jams.
The jams take the system from the unstable branch to the stable branch
at lower current.  Below $j_{max}$ the deterministic state is stable to perturbations.
One can view the maximum throughput state as a continuous
percolation transition for traffic jams.  A well known example of such a
transition out of an absorbing state is directed
percolation.  However, the critical behavior here is much
simpler than directed percolation; all of the critical exponents can
be obtained analytically based on a balance between cars going into a
jam and cars leaving the jam.  This leads to a random walk theory as
described later.

\begin{figure}
\vskip 2.5in
\caption{The fundamental diagram, $j(\rho)$.  Note the point of
maximum throughput is a phase transition where the order parameter
is the density of jammed cars.  The dotted line extending from
low density is the unstable high
density branch.}
\end{figure}

\section{Maximum Throughput Selection and Self-Organized Criticality}

We studied the model with a very specific boundary condition which
gives the highest throughput.  Maximum throughput, $j_{max}$, is
selected automatically when the left boundary condition is an
infinitely large jam and the right boundary is open.\cite{
nagel}$^,$\cite{ np95} Traffic which emerges from the megajam operates
precisely at highest efficiency.  This situation is shown in Fig. 2.
The horizontal axis is space and the vertical axis (down) is
increasing time.  The cars are shown as black dots which move to the
right.  The diagram allows us to follow the pattern in space and time
of the traffic.  Traffic jams show up as dense regions which drift to
the left, against the flow of traffic.  The structure on the left hand
side is the front of the megajam (cars inside the megajam are not
plotted).  Cars emerge from the big jam in a jerky way, before they
reach a smooth outgoing pattern operating at $j_{max}$.  Far away from
the front of the megajam all cars eventually reach maximum velocity.

\begin{figure}
\vskip 7in
\caption{Outflow from a dense region; only the front, or interface,
from the dense region is shown as the structure on
the left hand side (see text).  In the outflow region, a
phantom jam is triggered by a small disturbance. 
This is the structure on the right hand side.}
\end{figure}

We are now in a position to show that the outflow pattern is critical.\cite{ np95}
Far downstream from the megajam, a single car is perturbed slightly,
by reducing its velocity from $v_{max}$ to $v_{max} -1$.  This
particular car eventually accelerates to $v_{max}$.  In the meantime,
a following car may have come too close to the disturbed car and has
to slow down.  This initiates a chain reaction -- the phantom traffic
jam.  This is shown as the structure on the right hand side of the
figure, which was initiated near the top right hand corner.  Eventually,
the phantom jam always dissolves.  Sometimes the  phantom
jams are large and
sometimes they are small.  Any downstream car not moving at $v_{max}$
is considered to be part of the phantom jam.  In order to obtain
statistics for the properties of the noninteracting traffic jams, the
deterministic outflow is disturbed again, after the previous jam has
died out.  Fig. 3 shows the lifetime distribution of the jams.  This
is a power law, with exponent $3/2$.  
The absence of a characteristic
scale indicates that a critical state has been reached.

 No cataclysmic triggering event, like a traffic accident, is needed
to initiate large jams.  They arise from the same dynamical mechanism
as small jams and are a manifestation of the criticality of the
outflow regime.  Our natural intuition that large events come from
large disturbances is violated.  It does not make any sense to look
for reasons for the large jams.  The large jams are fractal, with
small sub-jams inside big jams ad infinitum.  Between the subjams are
``holes'' of all sizes where cars move at maximum velocity.  This
represents the irritating slow and go driving pattern that we are all
familiar with in congested traffic.  On the diagram, it is possible to
trace the individual cars and observe this intermittent pattern.  This
behavior gives rise to $1/f$ noise.
\begin{figure}
\vskip 3in
\caption{Lifetime distribution $P(t)$ for emergent jams in the outflow region;
average over more than 65\,000 avalanches.  The dotted
line has slope $3/2$.  All jams larger
than a numerically imposed cutoff at $t=10^6$ were removed from the
data  base.  This imposed cutoff can be made infinite, which indicates
that the outflow is precisely critical.}
\end{figure}

\subsection{Random Walk Theory}

It is, perhaps, surprising that the seemingly complicated structure
shown in Fig. 2 is described by such a simple apparent exponent.  The
lifetime exponent, $3/2$, appears to be the same as first return time
exponent for a one-dimensional random walk.  In fact, one can make a
self-consistent random walk theory which is valid as long as the jams,
themselves, are dense.  We will return to examine the density of jams
later.

Let us consider a single jam in a large system. 
When the vehicle at the front of the jam accelerates to
maximum velocity it leaves the jam forever.  The rate at which vehicles
leave the jam is determined by the probabilistic rules for
acceleration.  Vehicles, of course, can be added to the jam at the
back end.  These vehicles come in at a rate which depends on the
density and velocity of cars behind the jam.  
If the jam as a whole is dense, then one can
ignore internal branching mechanisms where a car may accelerate to maximum
velocity (thereby changing the number $n$ of jammed cars), but not
actually leave the jam without having to slow down again.  The density
of a jam is defined as the number of jammed vehicles, $n$, divided
by $w$, the distance between the leftmost and rightmost jammed vehicles.

The probability distribution, $P(n,t)$, for the number of cars
in the jam  at time, $t$, is determined by the following equation
for large $n$ and $t$:
\begin{equation}
 {\partial P \over \partial t} = (r_{out} - r_{in}){\partial P
\over \partial n} + {r_{out} +r_{in} \over 2}{\partial^2 P \over
\partial n^2} \quad .\label{continuum}
\end{equation}
The quantities $r_{in}$ and $r_{out}$ are phenomenological parameters
that depend on the density behind the jam and the rate at which cars leave
a jam.  They are independent of  the number of cars in the jam.

When the density behind the jam is such that the rate of cars entering
the jam is equal to the intrinsic rate that cars leave the jam, then
the first term on the right hand side of Eq. (1) vanishes, and the jam is
formally equivalent to an unbiased random walk in one
dimension,\cite{ Feller} or the diffusion equation.  The first return
time of the walk  corresponds to the lifetime of a phantom jam.
This leads immediately to the result $P(t) \sim t^{-3/2}$ for the
lifetime distribution, which agrees with the numerical observation.

This argument also shows that the outflow from a megajam is 
self-organized critical.  This can be seen  by noting
that the outflow from a megajam occurs at the same rate as the
outflow from a phantom jam created by a perturbation downstream from
the megajam.  One jam's inflow is the other jam's outflow, so that
$r_{in}=r_{out}$.  Since this is
true at all scales, the branching structure is a fractal.
This also shows that maximum throughput corresponds to
the percolative transition for the traffic jams.  Starting from random
initial conditions in a closed system, the current at long times is
determined by the outflow of the longest-lived jam in the system.
The random walk theory enables us to determine all the critical
exponents for the maximum throughput state, as explained in
Table I.  The agreement between
our random walk theory and simulations of the model
is very good, except for relations involving the
spatial extent $w$ of the jam.  Here internal dynamics generates logarithmic
corrections, as shown in the next section.  Not surprisingly, these logarithmic
corrections are also responsible for $1/f$ noise.
\begin{table}[p]
\caption{Critical exponents at maximum throughput. The variable $t$ is time
and the variable $\Delta \rho$ is distance from the critical density.}
\vspace{0.4cm}
\begin{center}
\begin{tabular}{|c|c||c|}
\hline
Physical Quantity &  Variable & Relation  \\
\hline
Number of Vehicles in Surviving Jams & $n(t)$&$n\sim t^{1/2}$\\
Size of the Jam & $s(t)$& $s \sim nt \sim t^{3/2}$\\
Lifetime Distribution of Jams & $P(t)$&$P(t) \sim t^{-3/2}$\\
Number of Jammed Vehicles in All Jams & ${\bar n}(t)$& ${\bar n} \sim t^0$ \\
Width of Surviving Jams &$w(t)$& $ w \sim t^{1/2}\ln (t)$ \\
Cutoff time & $t_{co}(\Delta \rho)$& $ t_{co} \sim (\Delta \rho)^{-2}$\\
Probability for Infinite Jam & $P_{\infty}(\Delta \rho)$ & $P_{\infty} 
\sim\Delta \rho$\\
Distribution of Hole Sizes &$P_h(x)$& $P_h(x) \sim x^{-2}$\\
Power Spectrum &$S(f)$& $S(f) \sim 1/f$\\
\hline
\end{tabular}
\end{center}
\end{table}

\section{Avalanches and $1/f$ Noise in Traffic}

A relationship between spatial fractal behavior and long-range
temporal correlations can be formally established as follows:
Consider Fig. 4, which is a space-time plot of a phantom traffic
jam where only the jammed cars are plotted.  The sites which are
not occupied with cars or which have cars moving at maximum velocity
are all considered to be ``empty'' vacuum sites.
 This space-time fractal,
or avalanche, has
the amusing property that its cuts in different directions are fractals
themselves.  If one makes a constant time cut, a snapshot of the jam
at some instant in time is
seen.  The jammed cars may comprise a fractal with dimension $d_f \leq 1$.
If so, then the intersection of the entire  jam with a constant
space cut, in the time direction perpendicular to the previous cut,
 will also be a fractal (see Figure).  In fact,
since the jams drift backwards, the fractal dimension of the time
cut is exactly the same as the fractal dimension of the space cut!~\cite{ np95}
It turns out that a temporal sequence of points with fractal
dimension $d_f \leq 1$ gives a power spectrum \cite{ mpb94}$^,$~\cite{ pmb96}
\begin{equation}
S(f) \sim {1\over f^{d_f}} \qquad .
\end{equation}
Thus, the problem of calculating the power spectrum exponent has been
reduced to the problem of calculating the spatial fractal dimension
of the jammed cars at a given instant in time.

In order to proceed, one more step is needed. 
Consider
a set of points with dimension $d_f \leq 1$ embedded on a one dimensional
line.  By definition, these points are separated by intervals of empty sites,
or holes.  The distribution of hole sizes may be a power law, i.e.
\begin{eqnarray}
P_h(x) \sim & x^{-\tau_h} \quad . \nonumber \\
 \quad{\rm If\ so, then }\quad d_f = & \tau_h -1 \quad ,
\end{eqnarray}
as long as $\tau_h \leq 2$.
In the next section, we calculate the characteristic exponent for the
distribution of hole sizes and find $\tau_h=2$ 
exactly.  This leads immediately
to the result that the power spectrum \cite{ np95}
\begin{equation}
S(f) \sim 1/f \qquad .
\end{equation}

Since $\tau_h =2$, $d_f=1$ and the jams are marginally dense.  Even
though the fractal dimension of jammed cars is unity, the jammed cars
have zero density inside a very large jam.  This is because the
average size of holes inside the jam is diverging logarithmically.
Thus we reach the fortuitous conclusion that the branching behavior of jams
leads to complicated intermittent dynamics with a $1/f$ power
spectrum, but the random walk theory for the jams, which ignores
branching, is still valid up to logarithmic corrections.
\begin{figure}[p]
\vskip 4.2in
\caption{ Schematic of a space-time plot of an emergent jam. 
 The horizontal direction
is space and the vertical direction is time, as in Fig. (2).
 Only vehicles with $v < v_{max}$, 
   are plotted.  The horizontal line is the constant
time cut of the system, where a spatial fractal would be observed, and
the vertical line is a constant space cut of the system, where $1/f$ noise
would be observed.}
\end{figure}

\subsection{A Cascade Equation for the Branching Jams}

We now analyze in detail the branching behavior of jams with $v_{max} >1$
\cite{ remark} in
terms of a phenomenological cascade equation.  A very large phantom
jam, at a fixed point in time, consists of small dense regions of
jammed cars, which we call subjams, separated by intervals, holes,
where all cars move at maximum velocity. We consider the
subjams to have size one.

Holes between the subjams
 are created at small scales by the probabilistic rules
for acceleration.  Each subjam can create small holes in front of it.
We will ignore the details of the injection mechanism, and assume that
there is a steady rate at which small holes are created in the interior
of a very long lived jam.  We also assume that the
 interior region of a long-lived jam reaches a steady state
distribution of hole sizes.  
We do not explicitly study the distribution
of hole sizes at small scales.

In order to determine the asymptotic scaling of the large holes
in the interior of a long-lived jam,
it is necessary to isolate the dominant mechanism in the cascade process for
large hole generation.  This mechanism is
the dissolution of one subjam.  When one subjam
dissolves because the cars in it accelerate to maximum velocity,
the two holes on either side of it merge to form one larger hole.  Holes at
any large scale are created and destroyed by this same process.
This mechanism links different large scales together, and we propose
that it gives the leading order contribution at large hole sizes.
In the steady state, the creation and destruction of large holes must
balance.  This leads to a cascade equation for holes of size $x$:
\begin{equation}
\sum_{u=x+1}^{\infty}<h(x)h(u-x)> = \sum_{x'=1}^{x-2}<h(x')h(x-x'-1)>
\quad . \label{cascade}
\end{equation}
Here, the angular brackets denote an ensemble average over all holes in the
jam, and the quantity $h(x)h(u-x)$ denotes a configuration where
a hole of size $x$ is adjacent to a hole of size $(u-x)$.  The right hand side
of this equation represents the rate at which holes of size $x$ are
created, and the left hand side represents the rate at which holes of
size $x$ are destroyed.

Now, we make an additional ansatz; namely, for
large $x$, 
\begin{equation}
<h(x')h(x-x'-1)>= G(x) \qquad , 
\end{equation}
independent of $x'$ to leading order.
That is, to leading order the probability to have two adjacent holes,
whose sizes sum to $x$ is independent of the size of either hole.
$G(x)$ then also scales the same as $P_h(x)$, the probability to have a hole
of size $x$.
Thus Eq.~\ref{cascade}, to leading order, can be written
\begin{equation}
\sum_{u=x}^{\infty} G(u) \sim xG(x)   \quad.
\end{equation}
Differentiating leads to
\begin{equation}
x{\partial G(x) \over \partial x} = -2G(x) \quad ; \quad G(x) \sim {1\over x^2}
\quad .
\end{equation}
Thus the distribution of hole sizes decays as 
\begin{equation}
P_h(x) \sim x^{-\tau_h} \quad ; \quad {\rm with \ \ } \tau_h =2 \quad .
\end{equation}
It is interesting to note that the cascade equation (\ref{cascade}) is
identical to the dominant mechanism in the exact cascade equation for
forests in the one-dimensional forest fire model.\cite{ Paczuski.93.Bak}
The exponent
$\tau_h =2$ is the same as the distribution
exponent for the  forests, which has been obtained exactly.\cite{ Drossel.1dff}
Since $\tau_h = d_f +1$,  $\tau_h < 2$ implies that
the equal time cut of the jam clusters is fractal, otherwise not.  The
point $\tau_h =2$ is the boundary between fractal and dense behavior.
At this special point, the random walk theory 
can still be expected to apply,
although with logarithmic corrections.   The borderline of fractality
gives precisely $1/f$ behavior.

\subsection{Numerical measurements}

We measured the distribution of hole sizes, as shown in Fig. 5.  
This was accomplished by running a phantom traffic jam until the cluster
reached a width of 8192, and storing the configuration of jammed cars
at that time.  About 60 configurations of the same size were used.
The observed slope agrees well with the prediction $\tau_h =2$.

The power spectrum, shown in Fig. 6,
 was measured in a closed system with a disturbance
rate $p_{free}= 0.00005$ in the steady state.  At low frequencies the
power spectrum is consistent with the prediction $S(f) \sim 1/f$
based on our microscopic theory for traffic.
\begin{figure}[p]
\vskip 3in
\caption{Probability distribution $P_h$ for hole-sizes $x$
for $v_{max}=2$.  The dotted
line has slope $-2$.  The average is over 60~configurations, which all
have width $w=2^{13}$.}
\end{figure}
\begin{figure}[p]
\vskip 3in
\caption{Power spectrum, $S(f)$, smoothed by averaging, for a closed
system of length~$L=10^5$, with $p_{free}=0.00005$ and $v_{max}=5$.
Dotted line has slope $-1$.}
\end{figure}

\section{Applications to real traffic}

The
following  results
should be general enough to be important for traffic:
\begin{itemize}

\item
The concept of critical phase transitions is helpful for
characterizing real traffic.  The most efficient state for traffic
that can be achieved is a critical state with jams of all sizes.  Open
systems will tend self-organize to this critical state.  Spontaneous
small fluctuations can lead to large emergent traffic jams.

\item
Technological advancements
such as cruise-control or radar-based driving support will tend to
reduce the fluctuations at
maximum speed similar to our limit, thus increasing the range of
validity of our results.  One unintended consequence of these flow control
technologies is that, if they work, they will in fact push the traffic
system closer to its underlying critical point, thereby making prediction,
planning, and control more difficult.

\item
The fact that traffic jams at  the border of fractal behavior
means that, from a single ``snapshot'' of a traffic system, one will
not be able to judge which traffic jams come from the same `reason'.
Concepts like queues  or single waves do not make
sense when traffic is close to criticality. Phantom traffic jams
emerge spontaneously from the dynamics of branching jam waves, and
give rise to $1/f$ noise.

\item
The regime near maximum throughput  corresponds to large
``holes'' operating practically at  $j_{max}$ and critical density, plus a
network of branched subjams.
The fluctuations found in the 5-minute-measurements of
traffic at capacity~\cite{ hall} therefore reflect the fact that
traffic flow is intermittent and
inhomogeneous with essentially two states (jammed and
maximum throughput).  The result of each 5-minute-measurement depends
on how many jam-branches are measured during this period.

\end{itemize}

\section{Traffic and Economics}

Traffic jams have negative economic impact.
One may note that in 1990 (1980), 14.8\% (16.4\%)
of the U.S. GNP was absorbed by passenger and
freight transportation costs.~\cite{ Eno}
The conventional view is that one should try to get rid of traffic jams
in order to increase efficiency and productivity.
However, we find that the critical state, with traffic jams of all
sizes, is the most efficient state that can actually be achieved.
A carefully prepared state where all cars move at maximum velocity
would have higher throughput, but it would be dreadfully unstable.
The very efficient state would catastrophically
collapse from any small fluctuation.  A similar situation occurs in the
familiar
sand pile models of SOC.\cite{ btw}  One can prepare a sand pile with
a supercritical slope, but that state is unstable to small perturbations.
Disturbing a
supercritical pile will cause a collapse of the entire system in one
 gigantic avalanche.

But there is perhaps even a deeper relationship between traffic and
economics (see also Refs. [15,16]).  In an economy, humans interact by
exchanging goods and services.  In the real world, each agent has
limited choices, and a limited capability to monitor his changing
environment.  This is referred to as bounded rationality.  The
situation of a car driver in traffic can be viewed as a simple example
of an agent trying to better his condition in an economy.  Each
driver's maximum speed is limited by the other cars on the road and
posted speed limits.  His distance to the car in front of him is
limited by his ability to stop and his need for safety in view of the
unpredictability of other drivers.  He is also exposed to random shocks
from the road or from his car.  He may be absent minded.  If traffic
is a paradigm for economics in general, then perhaps we have found a
new economic principle: the most efficient state that
can be achieved for an economy 
is a critical state with fluctuations of all sizes.

\section*{Acknowledgments}
This work was supported in part by the U.S. Department of Energy
Division of Materials Science, under contract DE-AC02-76CH00016. MP
thanks the U.S. Department of Energy Distinguished Postdoctoral
Research Program for financial support.  We thank Per Bak for helpful
comments on the manuscript.


\begin{thebibliography}{99}
\bibitem{ remark}
Note that for $v_{max}=1$ the jams do not branch, and
$1/f$ noise is not observed.
\end{thebibliography}
\end{document}